\documentclass[twocolumn]{aastex631}

\usepackage{ragged2e}
\usepackage{color}
\usepackage{xspace}
\usepackage[version=4]{mhchem} 
\usepackage[nameinlink]{cleveref}
\usepackage{longtable}

\newcommand\jwst{\em JWST}
\newcommand\hst{\em HST}

\newcommand\eureka{\texttt{Eureka!}\xspace}
\newcommand\firefly{\texttt{FIREFLy}\xspace}

\newcommand{\POSEIDON}{\texttt{POSEIDON}\xspace}
\newcommand\picaso{\texttt{PICASO}\xspace}

\newcommand\rfast{\texttt{rfast}\xspace}
\newcommand\tiberius{\texttt{Tiberius}\xspace}
\newcommand{\planetname}{GJ~486b\xspace}
\usepackage[symbol]{footmisc}

\DeclareSymbolFont{UPM}{U}{eur}{m}{n}
\DeclareMathSymbol{\umu}{0}{UPM}{"16}
\newcommand\micro{$\umu$}
\newcommand\microns{\micro m\xspace}

\newcommand\rsun{R$_{\odot}$}

\usepackage[version=4]{mhchem} 

\makeatletter
\newcommand*{\linktocite}[2]{%
  \hyper@natlinkstart{#1}#2\hyper@natlinkend}
\makeatother

\accepted{April 7, 2023}

\submitjournal{ApJL}

\shorttitle{GJ 486b with JWST NIRSpec/G395H}
\shortauthors{Moran \& Stevenson et al.}
\graphicspath{{./}{figs/}}

\begin{document}

\title{High Tide or Riptide on the Cosmic Shoreline? A Water-Rich Atmosphere or Stellar Contamination for the Warm Super-Earth GJ~486b from JWST Observations}

\author[0000-0002-6721-3284]{Sarah E. Moran}
\affiliation{Department of Planetary Sciences and Lunar and Planetary Laboratory, University of Arizona, Tuscon, AZ, USA}

\correspondingauthor{Sarah E. Moran \& Kevin B. Stevenson}
\email{semoran@lpl.arizona.edu, Kevin.Stevenson@jhuapl.edu}

\author[0000-0002-7352-7941]{Kevin B. Stevenson}
\affiliation{Johns Hopkins APL, Laurel, MD, 20723, USA}

\author[0000-0001-6050-7645]{David K. Sing}
\affil{Department of Earth \& Planetary Sciences, Johns Hopkins University, Baltimore, MD, USA}
\affil{Department of Physics \& Astronomy, Johns Hopkins University, Baltimore, MD, USA}

\author[0000-0003-4816-3469]{Ryan J. MacDonald}
\affil{Department of Astronomy, University of Michigan, 1085 S. University Ave., Ann Arbor, MI 48109, USA}
\affil{NHFP Sagan Fellow}

\author[0000-0002-4207-6615]{James Kirk}
\affil{Department of Physics, Imperial College London, Prince Consort Road, London, SW7 2AZ, UK}

\author[0000-0002-0746-1980]{Jacob Lustig-Yaeger}
\affil{Johns Hopkins APL, Laurel, MD, 20723, USA}

\author[0000-0002-1046-025X]{Sarah Peacock}
\affil{University of Maryland, Baltimore County, MD 21250, USA}
\affil{NASA Goddard Space Flight Center, Greenbelt, MD 20771, USA}

\author[0000-0002-4321-4581]{L. C. Mayorga}
\affil{Johns Hopkins APL, Laurel, MD, 20723, USA}

\author[0000-0002-9030-0132]{Katherine A. Bennett}
\affiliation{Department of Earth \& Planetary Sciences, Johns Hopkins University, Baltimore, MD, USA}

\author[0000-0003-3204-8183]{Mercedes L\'opez-Morales}
\affil{Center for Astrophysics ${\rm \mid}$ Harvard {\rm \&} Smithsonian, 60 Garden St, Cambridge, MA 02138, USA}

\author[0000-0002-2739-1465]{E. M. May}
\affiliation{Johns Hopkins APL, Laurel, MD, 20723, USA}

\author[0000-0003-4408-0463]{Zafar Rustamkulov}
\affiliation{Department of Earth \& Planetary Sciences, Johns Hopkins University, Baltimore, MD, USA}

\author[0000-0003-3305-6281]{Jeff A. Valenti}
\affil{Space Telescope Science Institute, Baltimore, MD 21218, USA}

\author[0000-0002-4489-3168]{J\'ea I. Adams Redai}
\affil{Center for Astrophysics ${\rm \mid}$ Harvard {\rm \&} Smithsonian, 60 Garden St, Cambridge, MA 02138, USA}

\author[0000-0003-4157-832X]{Munazza K. Alam}
\affil{Carnegie Earth \& Planets Laboratory, Washington, DC, 20015, USA}

\author[0000-0003-1240-6844]{Natasha E. Batalha}
\affil{NASA Ames Research Center, Moffett Field, CA, USA}

\author[0000-0002-3263-2251]{Guangwei Fu}
\affil{Department of Physics \& Astronomy, Johns Hopkins University, Baltimore, MD, USA}

\author[0000-0002-9032-8530]{Junellie Gonzalez-Quiles}
\affiliation{Department of Earth \& Planetary Sciences, Johns Hopkins University, Baltimore, MD, USA}

\author[0009-0009-3217-0403]{Alicia N. Highland}
\affil{Department of Astronomy, University of Michigan, 1085 S. University Ave., Ann Arbor, MI 48109, USA}

\author[0000-0002-0493-1342]{Ethan Kruse}
\affil{NASA Goddard Space Flight Center, Greenbelt, MD 20771, USA}
\affil{Department of Astronomy, University of Maryland, College Park, MD 20742.}
\affil{Center for Research and Exploration in Space Science and Technology, NASA/GSFC, Greenbelt, MD 20771}

\author[0000-0003-3667-8633]{Joshua D. Lothringer}
\affil{Department of Physics, Utah Valley University, Orem, UT, 84058 USA}

\author[0000-0003-3455-8814]{Kevin N. Ortiz Ceballos}
\affil{Center for Astrophysics ${\rm \mid}$ Harvard {\rm \&} Smithsonian, 60 Garden St, Cambridge, MA 02138, USA}

\author[0000-0001-7393-2368]{Kristin S. Sotzen}
\affiliation{Johns Hopkins APL, Laurel, MD, 20723, USA}
\affiliation{Department of Earth \& Planetary Sciences, Johns Hopkins University, Baltimore, MD, USA}

\author[0000-0003-4328-3867]{Hannah R. Wakeford}
\affil{School of Physics, HH Wills Physics Laboratory, University of Bristol, Bristol, UK}




\begin{abstract}

Planets orbiting M-dwarf stars are prime targets in the search for rocky exoplanet atmospheres. The small size of M dwarfs renders their planets exceptional targets for transmission spectroscopy, facilitating atmospheric characterization. However, it remains unknown whether their host stars' highly variable extreme-UV radiation environments allow atmospheres to persist. With {\jwst}, we have begun to determine whether or not the most favorable rocky worlds orbiting M dwarfs have detectable atmospheres. Here, we present a 2.8–-5.2\,$\micron$ {\jwst} NIRSpec/G395H transmission spectrum of the warm (700 K, 40.3$\times$ Earth's insolation) super-Earth \planetname (1.3 R$_\oplus$ and 3.0 M$_\oplus$). 
The measured spectrum from our two transits of \planetname deviates from a flat line at  $2.2-3.3\sigma$, based on three independent reductions. Through a combination of forward and retrieval models, we determine that \planetname either has a water-rich atmosphere (with the most stringent constraint on the retrieved water abundance of H$_2$O $>$ 10\% to 2$\sigma$) or the  transmission spectrum is contaminated by water present in cool unocculted starspots. We also find that the measured stellar spectrum is best fit by a stellar model with cool starspots and hot faculae. While both retrieval scenarios provide equal quality fits ($\chi^2_{\nu} = 1.0$) to our NIRSpec/G395H observations, shorter wavelength observations can break this degeneracy and reveal if \planetname sustains a water-rich atmosphere.

\end{abstract}

\keywords{JWST, Terrestrial Exoplanet Atmospheres, Transmission Spectroscopy}

\section{Introduction} \label{sec:intro}

Understanding the stability and longevity of atmospheres on rocky planets orbiting M dwarfs is paramount for understanding which, if any, of these planets may ultimately support life. However, given the high activity of most M-dwarf stars \citep[e.g.,][]{Peacock2019b}, their planets are subject to extreme-UV radiation regimes that may 
remove any significant atmosphere through escape processes \citep[e.g.,][]{Airapetian2020, Kasting1983,Zahnle2017,Airapetian2017}. This high activity also persists over much longer timescales given the long lifetimes of M dwarfs compared to larger stars \citep[e.g.,][]{Loyd2021}. M dwarfs also have the potential to impart spurious features into the transmission spectrum from inhomogenities in the stellar photosphere, a phenomenon called the ``Transit Light Source effect'' (TLS) \citep{Rackham2018}, also known as stellar contamination \citep{Apai2018,Barclay2021,Garcia2022,Barclay2023}.

Rocky worlds ($\leq$ 1.4R$_\oplus$) are not predicted to 
retain hydrogen/helium-dominated atmospheres \citep{Rogers2015,Rogers2021}. This 
has been confirmed by 
observations of %
terrestrial planets, 
including 
the TRAPPIST-1 planets \citep{deWit2016,deWit2018,Wakeford2019,Garcia2022,gressier2022}, GJ~1132b \citep{DiamondLowe2018,Mugnai2021,LibbyRoberts2022}, 
the L98-59 system \citep{Damiano2022,Zhou2023}, LTT~1445Ab \citep{DiamondLowe2022}
and LHS 3488b \citep{Kreidberg2019,DiamondLowe2020}. However, many of these 
observations do not preclude higher mean molecular weight secondary atmospheres for these small planets \citep{moran2018,Damiano2022}. 

 As part of the Cycle 1 {\jwst} General Observer (GO) Program 1981 (PIs: K. Stevenson \& J. Lustig-Yaeger), we are searching for atmospheric signatures on rocky planets around M dwarfs. Our program focuses reconnaissance on 
 carbon dioxide (CO$_2$) and methane (CH$_4$), believed to produce the strongest signals in terrestrial atmospheres \citep{KalteneggerTraub2009,Lustig-Yaeger2019b}. Both have 
 strong bands 
 between 3 and 5 $\mu$m, which can be 
 probed by 
 {\jwst}. 
 Secondary atmospheric CO$_2$ is also
 potentially common across a 
 range of terrestrial planetary conditions via outgassing 
 \citep{Lincowski2018}, as seen on Venus, Earth, and Mars. 
 Using {\jwst}, Program 1981 has already enabled 
 a strong constraint
 on Earth-sized exoplanet LHS~475b, ruling out Earth-like, hydrogen/helium, water, or methane-dominated clear atmospheres \linktocite{Lustig-YaegerFu2023}{(Lustig-Yaeger \& Fu et al.} \citeyear{Lustig-YaegerFu2023}).
 
 Our ultimate aim is to trace the proposed \textit{cosmic shoreline}, defined by \cite {Zahnle2017}. The cosmic shoreline describes the relationship between a planet's escape velocity (${\rm v_{esc}}$) and insolation (${\rm I}$). This ``shoreline'' divides rocky bodies with atmospheres from those without and
 is shaped by 
 various processes that cause atmospheric loss. In the solar system, this relationship follows  ${\rm I \propto v{_{esc}}^4}$, 
 suggesting that atmospheric escape mechanisms are dominated by thermal processes \citep{Zahnle2017}. Both thermal processes, such as Jeans escape and hydrodynamic escape, and non-thermal processes, encompassing photochemical escape and ion escape, 
 cause composition-dependent atmospheric loss. 
 These escape processes can be enhanced in 
 planets around active stars
 through UV flaring 
 or 
 stellar winds. Thus, to understand any putative cosmic shoreline in the solar system or beyond, it is important to determine not only how planet size, mass, and atmospheric composition affect a planet's ability to retain an atmosphere, but
 also the effect of the host star's activity. 
These varying factors can reveal the mechanisms dominating atmospheric escape on a given world \citep[e.g.,][]{Wordsworth2022,McIntyre2023}.

 Here we present the results of our JWST-GO-1981 program observations for 
 \planetname, 
a 1.3 R$_\oplus$ and  3.0 M$_\oplus$ planet \citep{Caballero2022}, with a zero Bond albedo equilibrium temperature of 700 K. \planetname has one of the highest transmission spectroscopy metrics \citep{kempton2018} of any known terrestrial exoplanet \citep{Trifonov2021}, making it a favorable target for study. 
The measured mass and radius indicate that \planetname is likely composed of a 
small metallic core, a deep silicate mantle, and 
a thin volatile upper layer \citep{Caballero2022}, 
which could be resistant to escape given the quiescent M3.5 V host star (0.339 \rsun, T$_{\rm{eff}}$ = 3291 K; \citealp{Caballero2022}). Recent high-resolution observations of \planetname show that the planet does not possess a clear 1$\times$ solar atmosphere dominated by hydrogen/helium to high confidence ($\geq 5\sigma$). These observations also suggest that a clear, pure water atmosphere could be ruled out to low significance ($\leq 3\sigma$) \citep{RiddenHarper2023}. We contextualize these observations in light of our own findings in Section \ref{sec:conclusion}.

\section{JWST Observations of \planetname} \label{observations}

We observed two transits of \planetname using the Near InfraRed Spectrograph \citep[NIRSpec;][]{NIRSpec2022, NIRSpec2022_Exoplanets} G395H instrument mode, covering wavelengths $2.87-5.14$ {\microns} at an average native spectral resolution $\mathcal{R}$ $\sim$ 2700.
The G395H grating is split over two detectors, NRS1 and NRS2, with a gap from 3.72 to 3.82 $\mu$m.
The first transit observation commenced on 25 December 2022 at 11:38 UTC and the second on 29 December 2022 at 21:15 UTC. Each observation lasted 3.53 hours, which 
covered the 1.01 hour transit duration and the required baseline. Both observations used the NIRSpec Bright Object Time Series (BOTS) mode with the NRSRAPID readout pattern, S1600A1 slit, and the SUB2048 subarray. For this bright target (K$_{mag}$ = 6.4), we used 3 groups per integration and obtained 3507 integrations per exposure. 

\section{NIRSpec G395H Data Reduction}

We reduced the data 
using three separate pipelines: \eureka \citep{Eureka2022}, \firefly \citep{Rustamkulov2022,Rustamkulov2023}, and \tiberius \citep{Kirk2018,Kirk2019,Kirk2021}. 
Each pipeline analysis is described below. Appendix \ref{sec:data} contains the updated system parameters obtained from each reduction. The three reductions showed a consistent offset in the measured transit depth for the Transit 1, NRS2 detector relative to the other three white light curve depths. 
We rule out astrophysical effects for this discrepancy and corrected it in each reduction as described in Appendix \ref{sec:offset}.

\subsection{\eureka}


We use a modified version of the 
\texttt{jwst} Stage 1 pipeline, starting 
from the {\em \_uncal.fits} files. 
We perform group-level background subtraction before determining the flux per integration.  For each group, we exclude the region within 9 pixels of the trace before computing and subtracting a median background value per pixel column.  We process the {\em \_rateints.fits} files through the regular \texttt{jwst} Stage 2 pipeline, skipping the flat fielding and absolute photometric calibration steps when our goal is to derive the planet's spectrum at later stages. 
Conversely, we include these steps when our goal is to compute the flux-calibrated stellar spectrum (see Section \ref{sec:spotty_star}).
Stage 3 of {\eureka} converts the time-series of 2D integrations into 1D spectra using optimal spectral extraction \citep{Horne1986} and an aperture within 5 pixels of the trace.  We flag bad pixels at numerous points within this stage using thresholds optimized to minimize scatter in the white light curves.

For the NRS1 detector, we extract the flux from 2.777 -- 3.717 {\microns} and split the light into 47 spectroscopic light curves, each 20 nm (0.02 {\microns}) in width.  For the NRS2 detector, we adopt the same resolution in extracting 67 spectroscopic light curves spanning 3.825 -- 5.165 {\microns}.
For each detector, we manually mask 9 pixel columns that exhibit significant scatter in their individual light curves. 
Doing so improves the quality of the spectroscopic light curves and yields more consistent transit depths.

\begin{figure*}[ht!]
    \centering
    \includegraphics[width=0.99\textwidth]{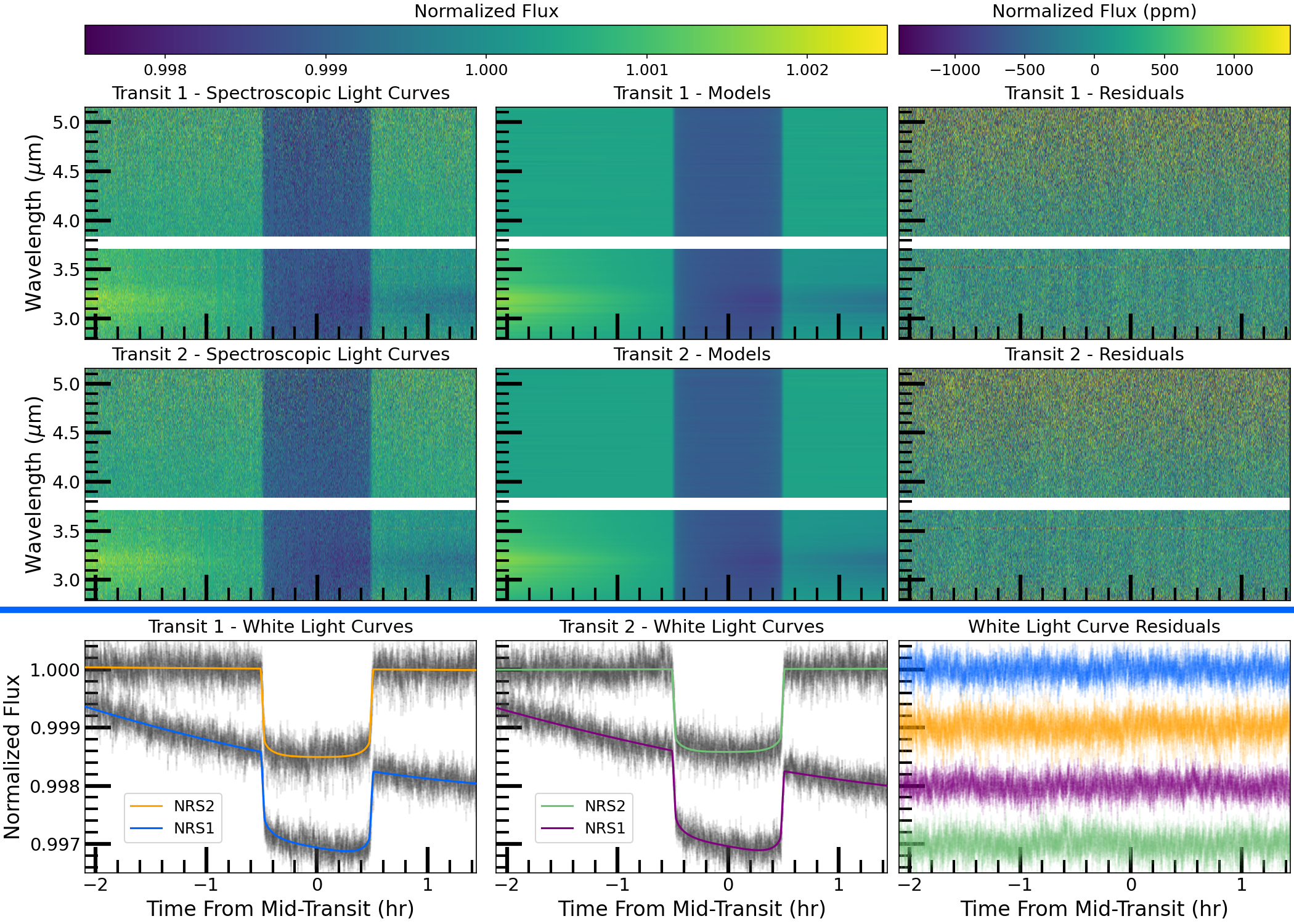}
    \caption{{\eureka} spectroscopic and white light curves from two transits of \planetname. The top two rows contain the spectroscopic light curves (left), our best-fit models (center), and subsequent residuals (right) for each transit.  Most evident in the data are wavelength-dependent ramps near 3.2 {\microns} that we readily remove. The bottom row depicts the white light curves from each detector (NRS1 and NRS2) before removing their systematic trends.  Correlated noise is evident in the residuals and is likely due to thermal cycling \citep{Rigby2022}.  The standard deviation of the normalized residuals is 140 ppm for NRS1 and 165 ppm for NRS2.
    The complete figure set (3 images, one for each reduction) is available in the online journal.}
    \label{fig:LCs}
\end{figure*}

With two NIRSpec detectors and two transit observations, we fit four white light curves and their systematics (see \Cref{fig:LCs}).  We determine the system parameters using \texttt{batman} \citep{batman2015} and fix the quadratic limb-darkening coefficients to those provided by \texttt{ExoTiC-LD} \citep{david_grant_2022_7437681}, assuming the stellar parameters given by \citet{Trifonov2021} and the MPS-ATLAS set 1 models \citep{Kostogryz2023}. For the NRS1 detector, we find that a quadratic trend in time provides the best fit.  For the NRS2 detector, a linear trend suffices to remove systematics.  \Cref{tab:Eureka_sys_params} lists our best-fit system parameters. 

When fitting the spectroscopic light curves (see \Cref{fig:LCs}), we fix the planet's transit midpoint, inclination, and semi-major axis to the weighted mean values in \Cref{tab:Eureka_sys_params}.  We fix the quadratic limb-darkening parameters 
to the values provided by \texttt{ExoTiC-LD} for each spectroscopic channel.
For the NRS1 detector, we also fix the quadratic term in our time-dependent systematic model to that of the best-fit white light curve value (Transit 1: $c_2=0.0335$, Transit 2: $c_2=0.0248$).  For all spectroscopic light curves, we fit for the zeroth and first-order terms ($c_0$ and $c_1$) of our polynomial.  Light curves from the NRS2 detector only require a linear model in time.  Including the term that rescales the uncertainties, each spectroscopic light curve has four free parameters, of which only the planet-to-star radius ratio is a physical parameter.

For each light curve, we first perform a least-squares minimization using the Powell method \citep{Powel1964} and then initialize our MCMC routine using our best-fit values.  We estimate the parameter uncertainties using \texttt{emcee} \citep{emcee2013} 
and, at each iteration, we increase the uncertainties by an average factor of $\sim$1.5 to achieve a reduced $\chi^2 = 1$.
All of our posteriors are Gaussian distributed and there are no parameter degeneracies.  

\subsection{\firefly}

We run the \texttt{jwst} pipeline through Stages 1 and 2 using the \texttt{uncal.fits} files. 
We utilize group-level 1/f subtraction and apply a scaled superbias to account for the vertical offset seen in NRS2 Transit 1. (See Section \ref{sec:offset}.) 
We correct for cosmic rays and bad and hot pixels in the Stage 2 output \texttt{rateints.fits} files and apply
a second 1/f correction at the integration level 
by masking the spectral trace 
and then calculating the median 
of the background pixels in each column. This value is then subtracted from the cleaned 2D image. 

We next cross-correlate each 2D image with the median aligned image to determine the x- and y-shifts of the spectral trace, which are used to align all 2D images. A Gaussian profile is then cross-correlated to each column in the y-direction and a fourth-order polynomial is fit in the x-direction to determine the spectral trace, which is used to extract the spectra. 


The white light curves for Transits 1 and 2 are fit from the extracted spectra by summing the spectra in the wavelength direction over a detector. We fit $a/R_{\star}$, limb darkening parameters, and the impact parameter \textit{b} using the weighted mean from both transits and both detectors. We then fix $a/R_{\star}$, \textit{b}, and the period, and fit for $R_P/R_{\star}$, $T_0$, and limb darkening in the white light curve. 
A low-order polynomial in time (third-order in NRS1 and up to fourth-order for NRS2) was used to model the baseline, with additional detrending parameters of the x- and y-shifts and superbias scale factor. 
We then fix the system parameters (presented in Table \ref{tab:firefly_sys_params}) and limb-darkening coefficients in each wavelength column to fit the spectroscopic light curves. 

\subsection{\tiberius}
With \tiberius 
we started by running STScI's \texttt{jwst} stage 0 pipeline on the uncal.fits files from the \texttt{group\_scale} step through \texttt{gain\_scale} step. We set \texttt{--odd\_even\_columns $=$ True} at the \texttt{ref\_pix} step and ran our own 1/f correction step at the group level prior to running \texttt{ramp\_fit}, which 
removes the median background flux for every column of every group's spectral image. We define the background 
as a 14-pixel-wide region that avoided 18 pixels centered on the curved trace, and mask 
bad pixels 
using our own custom bad pixel map. 
We subsequently ran \texttt{assign\_wcs} and \texttt{extract\_2d} to obtain the wavelength solution and
proceeded to run \texttt{Tiberius}'s spectral extraction on 
the \texttt{gainscalestep.fits}
files. 

First we oversampled each pixel by a factor of 10 using a linear interpolation. This allows us to measure the stellar flux at the sub-pixel level, which reduces noise in the light curves 
\citep{ERSFirstLook}. We used a fourth order polynomial to trace the NRS1 detector stellar spectrum 
and a sixth order polynomial for NRS2. We performed standard aperture photometry at every pixel column, with a 4-pixel-wide aperture. We performed an additional background subtraction step at this stage by calculating the background in 14 pixels on either side of the trace, excluding
7 pixels on each side. For NRS1 we fit these background pixels with a linear polynomial while for NRS2 we used a median since our defined background regions were mostly above the stellar trace.

We remove cosmic rays 
and 
residual bad pixels manually and then 
correct for small shifts in the stellar spectra along the dispersion direction by cross-correlating all spectra in the time-series with the first,
resampling each spectrum onto a common
pixel grid. 
Finally, we created a white light curve between 
2.75--3.72\,$\mu$m for NRS1 and 3.83--5.15\,$\mu$m for NRS2. Our spectroscopic light curves were created at 1 pixel resolution over the same wavelength range.


We 
fit the four white light curves (2 transits $\times$ 2 detectors) with \texttt{batman} \citep{Kreidberg2015}, leaving  
$a/R_*$, 
$R_P/R_*$, the orbital inclination ($i$), and the time of mid-transit ($T_0$) as free parameters, and fixing  
the period to the value from \cite{Trifonov2021}. 
For our white and spectroscopic light curves, we assumed quadratic limb darkening with coefficients fixed to values from 3D stellar atmosphere models \citep{magic2015stagger} using \texttt{ExoTiC-LD} \citep{david_grant_2022_7437681}. We adopted $T_{\mathrm{eff}} = 3340$\,K, [Fe/H] $=0.070$ and $\log g_* = 4.9155$ \citep{Trifonov2021}. 
For our systematics model we used a combination of polynomials: quadratic-in-time, linear-in-x-position, and linear-in-y-position, resulting in 
9 free parameters: 4 transit model parameters and 5 systematics model parameters. 

To determine the best fitting values and uncertainties, we used \texttt{emcee} \citep{emcee2013} with 90 walkers for two runs of 20,000 steps. After the first run we inflated our photometric uncertainties to give a reduced $\chi^2 = 1$ for our best-fitting model before the second run. Table \ref{tab:Tiberius_sys_params} summarizes the results of our 
white light curve fits. 
For our spectroscopic light curve fits, we fixed $a/R_*$, $i$ and $T_0$ to the weighted mean values from our 4 white light curve fits and 
only fitted for $R_P/R_*$ and the 
5 parameters defining our systematics model. 
Here we used a Levenberg-Marquadt sampler for computational speed as we had to fit 6876 spectroscopic light curves.

\section{Interpretation of \planetname's Transmission Spectrum} \label{sec:results}

The three 
data reductions produce 
consistent spectra with a slight slope on the blue end 
($\leq$ 3.7 $\mu$m) but are otherwise featureless. Here, we first quantify the significance of this slope in \planetname's spectrum. We then proceed to 
offer physical explanations of the spectrum through 
forward modeling and retrieval analyses.

\subsection{A Non-Flat Spectrum}

We performed a flat line hypothesis rejection test to determine the statistical significance of the 
slope in the 
transmission spectrum. We fitted the spectrum from each pipeline 
using two models: a flat featureless model that uses one free parameter for the transit depth, and a Gaussian spectral feature model with four free parameters: the flat transit depth and the central wavelength, amplitude, and width of a Gaussian feature added to the baseline featureless spectrum. 
We fitted both models to each dataset using the \texttt{dynesty} nested sampling code for Bayesian inference \citep{Speagle2020} and then used the Bayesian evidence to calculate the Bayes factor of each 
model \citep[e.g.,][]{Trotta2008,Trotta2017}. We then converted the Bayes factors to more classical 
``sigma'' detection significances using the relationship detailed by 
\citet{Benneke2013}.  

\begin{figure*}[ht!]
    \centering
    \includegraphics[width=\textwidth]{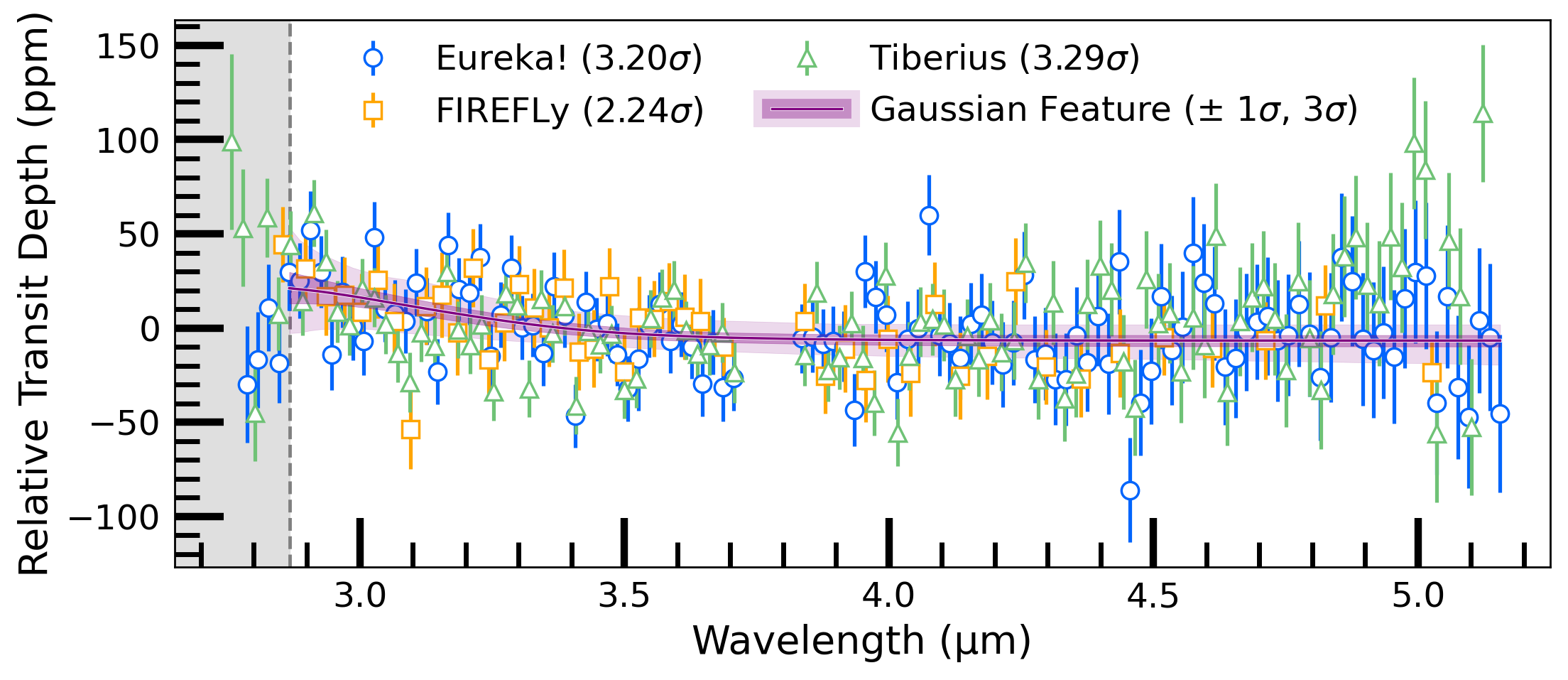}
    \caption{\label{fig:gaussian_spectra}\small 
     Relative transmission spectra of the three data reductions (\eureka: blue circles, \firefly: orange squares, \tiberius: green triangles). The median fit to the \eureka dataset using an agnostic Gaussian model is shown in purple bounded by $1 \sigma$ and $3 \sigma$ Bayesian credibility envelopes. The legend displays the statistical significance with which each reduction rules out a flat line in favor of the Gaussian model. Analyses of all three reductions reveal an uptick at the blue end of the wavelength range.  Instrument throughput deteriorates in the grey shaded region and the measured transit depths become unreliable; thus we exclude points within this region from our hypothesis rejection tests.  
     }
\end{figure*}

\Cref{fig:gaussian_spectra} 
demonstrates that each spectrum separately 
favors the Gaussian model and rejects a featureless spectrum. The strength of the signal detection is 
$3.20 \sigma$ for \eureka , 
$2.24 \sigma$ for \firefly , and 
$3.29 \sigma$ for \tiberius. The \firefly detection significance is 
lower due to slightly larger uncertainties associated with 
that reduction, which 
stem from \firefly's choice of spectroscopic binning 
to produce similar transit depth errors across the full wavelength range and wavelength-dependent baseline functions. Nevertheless, the same shape is seen in the spectra from the three pipelines.
Thus, the flat line hypothesis is rejected by all three analyses with varying confidence. Each individual reduction hypothesis rejection test is available in the online journal.

\vspace{10pt}
\subsection{Forward Modeling Tentatively Supports an Atmosphere with Water Vapor} \label{sec:forward_models}

We ran a suite of forward models 
using the stellar and planet parameters from \citet{Caballero2022}
to compare to each transmission spectrum. We also generated forward models using an updated stellar log(g) = $4.91\pm0.02$, obtained from our updated $a/R_s$ constraints (See Appendix \ref{sec:data}) \citep{Seager2003,sandfordkipping2017}, finding consistent results.



We 
focus on higher mean molecular weight scenarios to explain the transmission spectrum. 
For completeness, however, we simulate a 1000$\times$ solar metallicity atmosphere with a parameterized pressure-temperature profile in thermochemical equilibrium with \texttt{CHIMERA} \citep{Line2013b,Line2014-C/O} as in our previous work \citep{Lustig-YaegerFu2023}. We include the species \ce{H2O}, \ce{CH4}, CO, \ce{CO2}, \ce{NH3}, \ce{HCN}, \ce{H2S}, \ce{H2}, and He. The \texttt{CHIMERA} thermochemical equilibrium abundances result in a model spectrum that is primarily shaped by methane, carbon dioxide, and water. After generating the temperature-pressure profile and atmospheric abundances with \texttt{CHIMERA}, we use the radiative transfer suite of \picaso \citep{Batalha2019}, with opacities resampled to $R=10,000$ from \citet{Batalha2020}, to generate model spectra.

In each case, we bin the resulting model transmission spectrum to the resolution of the data before performing a reduced-$\chi^2$ comparison. The full datasets of all three reductions to which we fit our forward models and retrievals can be found with the Supplemental Materials. As with the Gaussian hypothesis tests, we exclude the data points in the grey shaded region of \Cref{fig:gaussian_spectra} from our model-fitting due to steeply-falling instrument throughput at these wavelengths ($<$ 2.87). 

\begin{figure*}
    \centering
    \includegraphics[width=\textwidth]{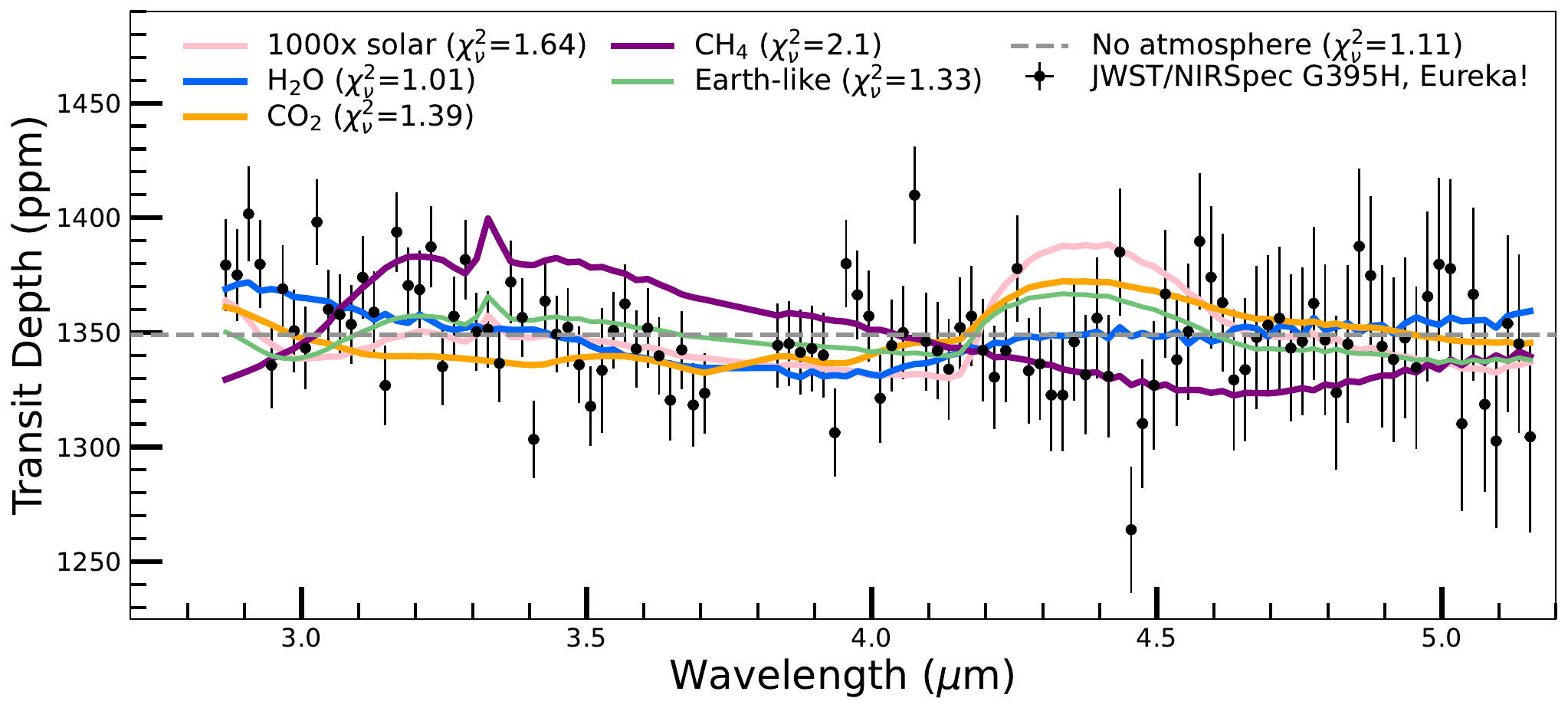}
\caption{Our final \eureka spectra of \planetname binned to R$\sim$200 (black points) compared to a set of {\picaso} forward models (colored lines: 1000$\times$ solar, pink; H$_2$O, blue; CO$_2$, orange; CH$_4$, purple; Earth-composition, green). A 1 bar, pure water atmosphere on \planetname fits the data with the lowest reduced-$\chi^2$ (1.01), and  a flat-line model (dashed grey line) is nearly as well fit by the data (reduced-$\chi^2$ = 1.11), though is weakly rejected by Gaussian vs flat line tests. Alternatively, stellar contamination with water in the atmosphere of the star, rather than the planet, can explain the observed transit depths (see Fig. \ref{fig:retrieval}).}
\label{fig:forward_models}
\end{figure*}

As shown in \Cref{fig:forward_models}, the slight slope and flatness of the spectra from each reduction allow us to confidently disregard low mean molecular weight atmospheres dominated by hydrogen/helium -- up to metallicities of 1000$\times$ solar -- to greater than 3$\sigma$. This improves upon the previous high resolution data obtained by \cite{RiddenHarper2023} that could only strongly rule out atmospheres up to a few times solar. Our 1000$\times$ solar metallicity atmosphere has an average mean molecular weight of 13.86 g/mol compared to the high resolution's 5 g/mol limit, though our constraint is less stringent for non-chemically consistent atmospheres (see Section \ref{sec:retrieval_main}).

We also compare the data from each reduction to a set of end-member forward models from \picaso with single-gas 1 bar, isothermal atmospheres. For ease of interpretation, we focus here on the results from the \eureka reduction, as we determined that it was the most representative dataset, with the smallest weighted average deviation from the median of all three reductions. However, the trend in best-fit agrees among all three reductions (for a complete description of each reduction's fit, see Table \ref{tab:models} in Appendix \ref{sec:forward_and_retrieval}). The slight slope on the blue end of NRS1 results in best-fitting (reduced-$\chi^2$ = 1.01) forward models that contain pure water vapor, as this molecule has a strong absorption feature from 2.2 to 3.7 $\mu$m, consistent with the slope we observe in NRS1.

Our data across all reductions also moderately to weakly rule out carbon-rich atmospheres of either CH$_4$ or CO$_2$ to 6.5$\sigma$ and 2.3$\sigma$, respectively. A flat-line model, representative of an airless body or a high-altitude (0.1 $\mu$bar) cloud deck, fit the data with reduced-$\chi^2$ = 1.11, which is statistically equivalent to the clear water atmosphere model within the forward modeling framework. However, between its equilibrium temperature and size, \planetname is not expected to support clouds to such low pressures, as there are few condensible species in this temperature range. Photochemical hazes could dampen the presence of any spectral features with a haze layer at this altitude and create a flat line spectrum \citep{gao2020,pidhorodetska2021,Caballero2022}; however, given the Bayesian evidence of the Gaussian absorption tests discussed above, the water atmosphere is the preferred explanation from the \picaso analysis for all reductions. We note that the \firefly reduction only weakly rejects the flat line hypothesis and, therefore, an airless planet or very hazy planet is still a possibility. In Figure \ref{fig:forward_models}, we show the results of our \picaso forward modeling compared to the \eureka data. The full set of results for each reduction is available in the online journal. 



\subsection{Retrievals Suggest a Water-rich Atmosphere or Unocculted Starspot Contamination} \label{sec:retrieval_main}

In addition to our forward model comparisons, we performed an atmospheric retrieval analysis to assess the robustness of our tentative evidence for a water-rich atmosphere and consider alternative astrophysical explanations. We apply two independent retrieval codes --- \POSEIDON \citep{MacDonald2017,MacDonald2023} and \rfast \citep{RobinsonArnaud2023} --- to all three data reductions to ensure reliable inferences. 


\subsubsection{Water-rich Atmosphere Scenario}  \label{sec:retrieval_water_rich}

Our \POSEIDON atmospheric retrieval considers six potential gases that can range in abundance from being trace volatiles to the dominant background gas: N$_2$, H$_2$, H$_2$O, CH$_4$, CO$_2$, and CO. The opacity contributions from these gases include line opacity \citep{Polyansky2018,Yurchenko2017,Tashkun2011,Li2015} and collision-induced absorption (CIA) from H$_2$-H$_2$, H$_2$-N$_2$, H$_2$-CH$_4$, H$_2$-CO$_2$, CO$_2$-CO$_2$, CO$_2$-CH$_4$, and N$_2$-N$_2$ \citep{Karman2019}. Since the mixing ratios must sum to unity, we have five free parameters describing their mixing ratios that each follow centered log-ratio (CLR) priors, ranging from $10^{-12}$ to 1, as described by \linktocite{Lustig-YaegerFu2023}{Lustig-Yaeger \& Fu et al.} (\citeyear{Lustig-YaegerFu2023}). The other free parameters are the isothermal temperature ($\mathcal{U}$ [200\,K, 900\,K]), the atmosphere radius at the 1\,bar reference pressure ($\mathcal{U}$ [0.9\,$R_{\rm{p, \, obs}}$, 1.1\,$R_{\rm{p, \, obs}}$]), and the log-pressure of an opaque surface ($\mathcal{U}$ [-7, 2], in bar). We calculate transmission spectra via opacity sampling at a resolving power of $R =$ 20,000 from 0.5--5.4\,$\micron$, with the lower wavelength limit set far below our shortest wavelength (2.8\,$\micron$) to later demonstrate how retrieval solutions diverge at optical wavelengths. These 8-parameter \POSEIDON retrievals used the \texttt{PyMultiNest} \citep{Feroz2009,Buchner2014} package to explore the parameter space with 2,000 live points.






\begin{figure*}[ht!]
    \centering
    \includegraphics[width=\textwidth]{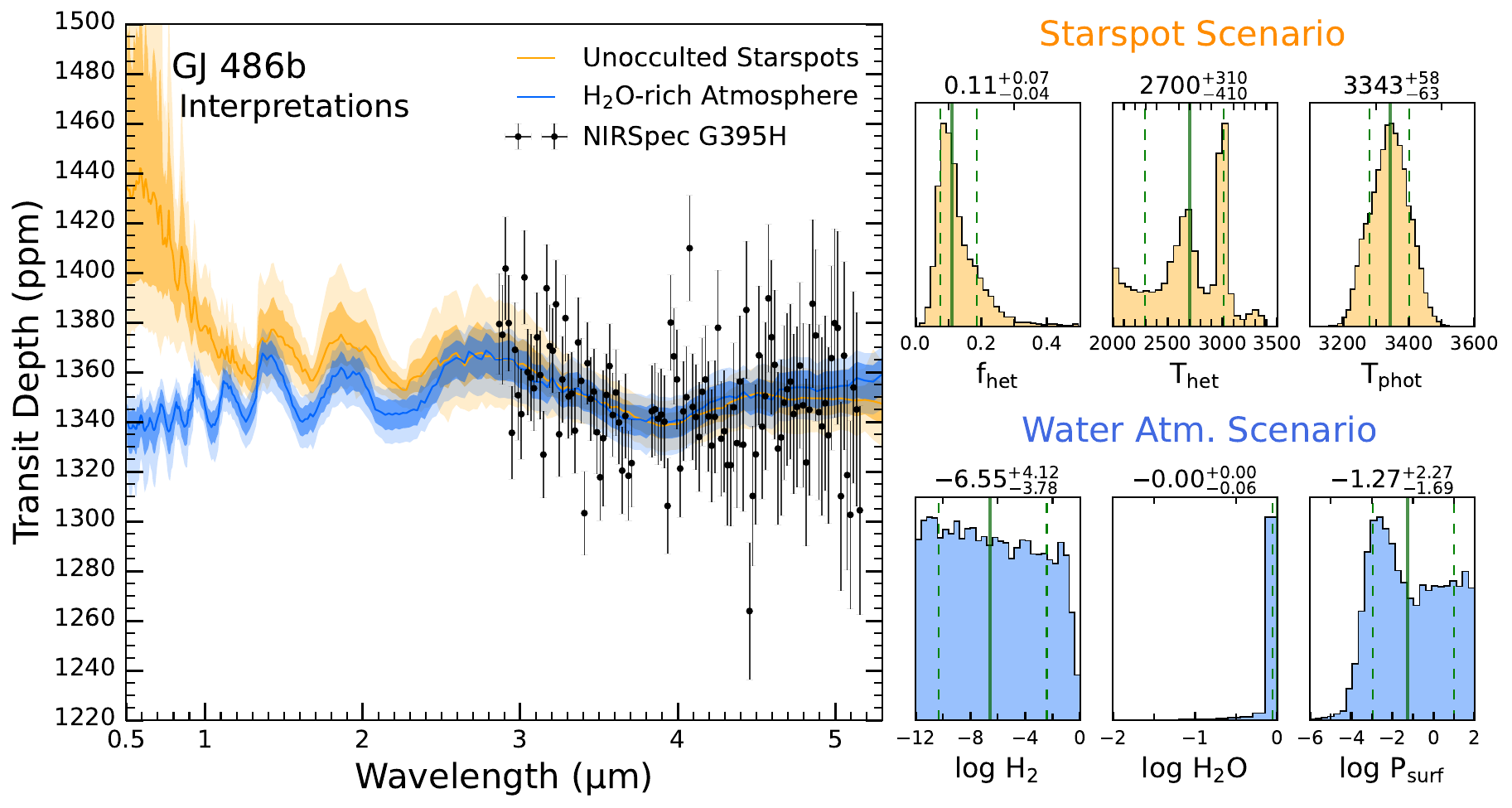}
    \caption{\label{fig:cosmic}\small
     \POSEIDON retrieval results for \planetname's transmission spectrum. Left: retrieved transmission spectra for two models compared to the {\jwst} NIRSpec G395H data from the \eureka reduction (black points with error bars). Two scenarios can equivalently explain \planetname's transmission spectrum ($\chi^2_{\nu} = 1.0$): unocculted starspots with no planetary atmosphere (orange contours) or a water-rich atmosphere with no starspots (blue contours). The median retrieved spectrum (solid lines) and 1$\sigma$ and 2$\sigma$ confidence intervals (dark and light contours) for each scenario are overlaid. Top right: posterior histograms for the unocculted starspot model, defined by the fractional coverage area of cold stellar heterogeneities/spots (f$_{het}$), the temperature of the heterogeneities/spots (T$_{het}$), and the stellar photospheric temperature (T$_{phot}$). Bottom right: posterior histogram for the water-rich atmosphere scenario, highlighting hydrogen and water's retrieved mixing ratios alongside the atmospheric surface pressure. Water is necessary to explain \planetname's spectrum, but the retrievals cannot differentiate between a water-rich planetary atmosphere or water contained in cool starspots that contaminate the transmission spectrum. The complete figure set (3 images, one for each reduction) is available in the online journal. 
     }
     \label{fig:retrieval}
\end{figure*}

\Cref{fig:retrieval} shows our \POSEIDON retrieval results for this atmospheric model scenario (blue retrieved spectrum and histograms) for the \eureka data reduction -- see the online figure set for the other two reductions. For \eureka and \firefly, the preferred explanation for the observed rise in the blue wavelengths of the transmission spectrum is H$_2$O opacity from the wing of the band centered on 2.8\,{\microns}. Bayesian model comparisons favor the presence of H$_2$O with Bayes factors of 133 and 8 (3.6$\sigma$ and 2.6$\sigma$) for \eureka and \firefly, respectively. The retrieved H$_2$O abundance posterior indicates that water is the most likely background gas (e.g., \eureka requires a H$_2$O mixing ratio $>$ 10\% to 2$\sigma$ confidence), with an upper limit ruling out a H$_2$-dominated atmosphere. The \eureka and \tiberius reductions also yield upper limits on the CH$_4$ and CO$_2$ abundances (see the Appendix, Figure~\ref{fig:POSEIDON_summary}). The \tiberius reduction, however, does not uniquely infer a water-rich atmosphere. Though a water-rich atmosphere remains the preferred solution for \tiberius, a secondary mode permits a clear, H$_2$-dominated atmosphere with no other gases contributing to the spectrum. This secondary mode reflects a solution where the wavelength dependence of H$_2$-H$_2$ CIA is used to fit the spectrum. This solution is unphysical since an H$_2$-dominated atmosphere will always contain other trace molecules with more prominent absorption features at these wavelengths. Upon further investigation, we found that the unphysical solution is driven by the upwards rise at the longest wavelengths that are only present in the \tiberius reduction (see \Cref{fig:gaussian_spectra}). We, therefore, conclude that a consistent explanation for \planetname's transmission spectrum, assuming the observed non-flatness is caused by atmospheric absorption, can be readily explained ($\chi^2_\nu \approx 1.0$) by a water-rich atmosphere --- in agreement with the forward models in Section~\ref{sec:forward_models}.

We also conducted single-composition atmospheric retrievals with \rfast for all three reductions. These retrievals consider atmospheres with a single absorbing gas alongside a spectrally inactive background gas with an agnostic mean molecular weight. Our \rfast retrieval model has 6 free parameters: the log-gas mixing ratio, $\log_{10} f_\mathrm{gas}$ ($\mathcal{U}$ [-12, 0]), the log-surface pressure, $\log_{10} P_0$ ($\mathcal{U}$ [-1, 6], in Pa), the surface temperature, $T_0$ ($\mathcal{U}$ [300, 1100]\,K), the mean molecular weight of the background gas, $m_b$ ($\mathcal{U}$ [2, 50]\,amu), the planet radius, $R_p$ ($\mathcal{U}$ [1.1, 1.4]\,$R_\oplus$), and the planet mass, $M_p$ ($\mathcal{N}$ [2.28, 0.12]\,$M_\oplus$). For the single gases, we consider, in separate retrievals, H$_2$O, CO$_2$, CO, and CH$_4$. The \rfast retrievals use \texttt{emcee} \citep{emcee2013} with 100 walkers for 15,000 steps, where the first 5,000 are discarded for burn-in.

We show our \rfast 1D posteriors in the Appendix (\Cref{fig:rfast_summary}). Our \rfast retrievals also identify a H$_2$O-rich atmosphere as a consistent explanation for the \eureka and \firefly reductions (though the lower limits on H$_2$O are weaker compared with \POSEIDON due to the combination of a free mean molecular weight, planet mass, and log-uniform vs. CLR priors). \rfast also finds that the \tiberius reduction permits lower mean-molecular weight atmospheres for similar reasons to \POSEIDON.

\subsubsection{Unocculted Starspot Scenario}  \label{sec:retrieval_starspots}

We now consider the potential for \planetname's host star alone to explain our observed transmission spectrum. Stellar heterogeneities (starspots and/or faculae) that are not occulted during transit can induce wavelength-dependent features in transmission spectra if the stellar intensity illuminating the planetary atmosphere differs from the overall average stellar intensity --- also known as the transit light source effect (TLS) \citep[e.g.,][]{Rackham2018}. This confounding stellar influence is a crucial consideration for transmission spectra of planets orbiting cool M dwarfs, such as GJ~486, since H$_2$O existing in cold starspots could mimic atmospheric signatures.

We implement stellar contamination retrievals with \POSEIDON following a similar approach to \citet{Rathcke2021}, based on the parameterization from \citet{Pinhas2018}. The contamination model is defined by four parameters: the stellar heterogeneity temperature, $T_{\rm{het}}$ ($\mathcal{U}$ [2300\,K, 1.2\,$T_{*, \rm{eff}}$]), the heterogeneity coverage fraction, $f_{\rm{het}}$ ($\mathcal{U}$ [0, 0.5]), the stellar photosphere temperature, $T_{\rm{phot}}$ ($\mathcal{N}$ [$T_{*, \rm{eff}}$, $\sigma_{T_{*, \rm{eff}}}$]), and the planetary radius, $R_p$ ($\mathcal{U}$ [0.9\,$R_{\rm{p, \, obs}}$, 1.1\,$R_{\rm{p, \, obs}}$]). For the priors, we adopt literature values of $T_{*, \rm{eff}} = 3340$\,K and $\sigma_{T_{*, \rm{eff}}} = 54$\,K \citep{Trifonov2021}. We calculate the stellar contamination factor by interpolating the \citet{Allard2012} grid of stellar PHOENIX models using the \texttt{pysynphot} package \citep{STScIDevelopmentTeam2013}.

\Cref{fig:retrieval} demonstrates that contamination from unocculted starspots, with no planetary atmosphere, provides an equally plausible ($\chi^2_{\nu} \approx 1.0$) alternative explanation to \planetname's transmission spectrum. In this scenario, the observed slope in the spectrum is still caused by the wing of an H$_2$O band, but the water resides in the host star. The \POSEIDON retrievals for all three data reductions yield a spot coverage fraction of $\sim 10\%$, but with relatively weak and inconsistent constraints on the spot temperature. Compared to a flat spectrum, the unocculted starspot model is preferred with Bayes factors of 255, 16, and 114 (3.8$\sigma$, 2.9$\sigma$, and 3.5$\sigma$) for \eureka, \firefly, and \tiberius, respectively. We stress that, while our present observations cannot distinguish between the water-rich atmosphere scenario and unocculted starspots, these two scenarios deviate substantially at shorter wavelengths (see \Cref{fig:retrieval}). Consequently, even in the case of aerosol-laden atmospheres \citep{rackham2022}, future observations at shorter wavelengths can readily distinguish which scenario is correct.

\subsection{A Spotty Star Best Explains the Stellar Spectrum} \label{sec:spotty_star}

To further investigate the possibility of stellar contamination, we return to the {\jwst}/NIRSpec G395H data to probe the Stage 3 stellar spectra and examine whether the star is consistent with a particular stellar model. Upon completing Stage 2 of the \texttt{jwst} pipeline with the flat fielding and absolute photometric calibration steps enabled, we noticed that only the region within 8 pixels of the trace is converted to units of MJy.  The remaining pixel regions are in DN/s, so we manually mask them before running Stage 3 of \eureka.  Due to the lack of unmasked background pixels, we disable Stage 3 background subtraction for this flux-calibrated reduction.  This change does not skew the final calibrated spectrum since we already performed group-level background subtraction in Stage 1.

\begin{figure*}[ht!]
    \centering
    \includegraphics[width=\textwidth]{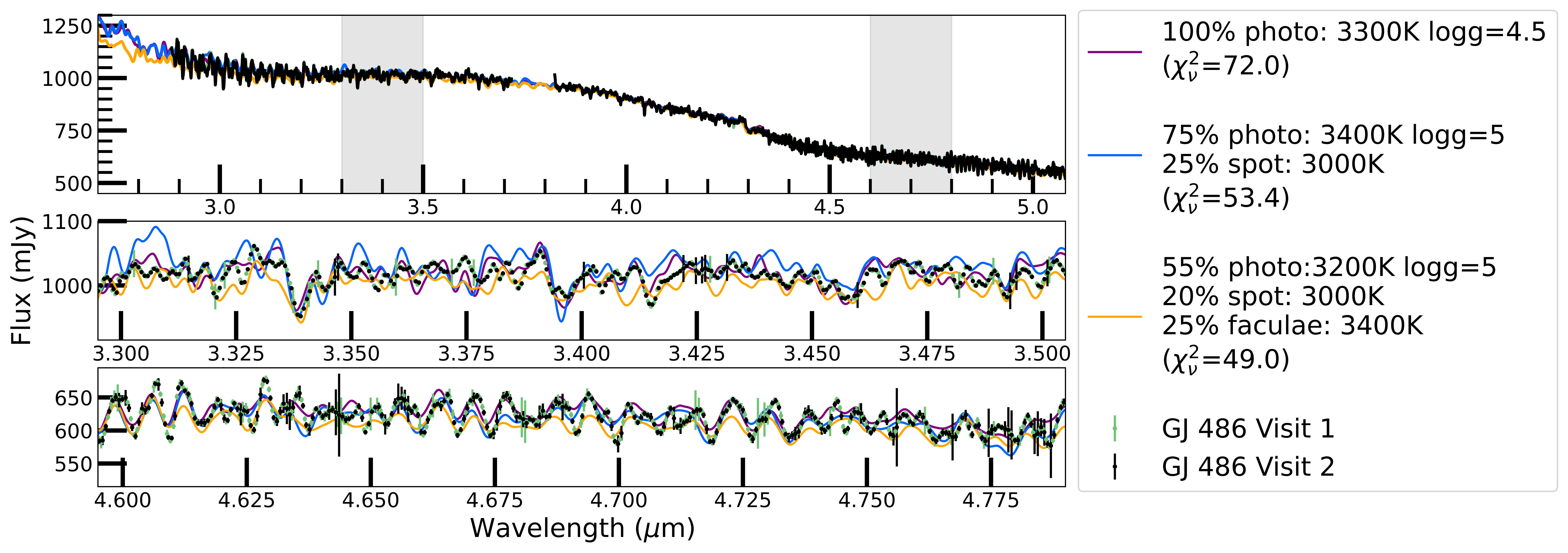}
    \caption{\small
     Best matching one-, two-, and three-component \texttt{PHOENIX} models to the Baseline GJ 486 spectra from Transits 1 (green) and 2 (black). The bottom two panels zoom-in on the grey highlighted regions of the top panel spectrum. When considering a one-component photosphere, a $T_{\rm{eff}}$ = 3300 K, log($g$)=4.5 cgs model is preferred (purple, $\chi^{2}_{\nu}$ = 72.0). When allowing for spots in a two-component model, a warmer $T_{\rm{eff}}$ = 3400 K, log($g$)=5 cgs photosphere with 25\% coverage of $T_{\rm{eff}}$ = 3000 K, log($g$)=5 cgs spots is the preferred model (blue, $\chi{^2}_{\nu}$ = 53.4). The best overall match to the observations is produced with a three-component photosphere+spots+faculae model that has a background photosphere with $T_{\rm{eff}}$ = 3200 K, log($g$)=5 cgs, 20\% spot coverage ($T_{\rm{eff}}$ = 3000 K, log($g$)=5 cgs), and 25\% faculae coverage ($T_{\rm{eff}}$ = 3400 K, log($g$)=5 cgs) (orange, $\chi^{2}_{\nu}$ = 49.0).}
     \label{fig:stellar}
\end{figure*}

To compute the stellar baseline spectrum, we exclude 1040 integrations during transit (1560 - 2599) and then compute median values along the time axis.  We manually mask a few obvious outliers before estimating the baseline spectrum uncertainties by computing the standard deviation in flux along the time axis.  Typical uncertainties are 3 -- 5 mJy, but can be as large as 55 mJy for some spectral channels.  The typical uncertainty values are consistent with the uncertainties derived from our standard spectral extraction routine.  We do not use the standard error calculation for our uncertainties. That is, we do not divide our uncertainties by the square root of the number of integrations because, as demonstrated below, the standard deviation in flux better represents the true uncertainty in our flux-calibrated spectrum.  We note that the derived baseline spectrum is remarkably consistent between both transits (see \Cref{fig:stellar}).

We used \texttt{PHOENIX} stellar models produced by \cite{Allard2012} to analyze whether the observed stellar baseline spectrum is best explained by a spotless or spotted star. We utilized the \cite{Allard2012} models, as in Section~\ref{sec:retrieval_starspots}, because they account for the formation of molecular bands including H$_2$O, CH$_4$, and TiO$_2$ and have higher ($\Delta\lambda$=2 \AA) resolution than the observations. This grid of models also has sufficient temperature and gravity coverage to model the photospheres of M-dwarf stars and their spots and faculae ($T_{\rm eff}$ $\geq$ 2000 K, log($g$) = 0 -- 6 cm s$^{-2}$).

We employed single \texttt{PHOENIX} models to represent spotless (or one-component) stars. We used weighted linear combinations of \texttt{PHOENIX} models to create inhomogeneous models. Two-component models include one model with $T_{\rm eff}$ $\geq$ 3000 K to represent the background photosphere and a second, cooler model with $T_{\rm eff}$ $\leq$ $T_{\rm eff, photosphere}$ - 100 K to represent spots. Three-component models include an additional $T_{\rm eff}$ $\geq$ $T_{\rm eff, photosphere}$ + 100 K model to represent faculae. In the two- and three-component models, all spots have the same $T_{\rm eff}$ and log($g$), as do the faculae. Linear combinations were computed by interpolating the spot and faculae models onto the photosphere wavelength grid before summing the fluxes in a weighted fraction where the photosphere was required to be $\geq$ 50\% of the total.

To compare the models to the observed baseline spectra, we converted the native wavelengths from \AA\ to $\mu$m and the flux densities from ergs s$^{-1}$ cm$^{-2}$ cm$^{-1}$ to fluxes in units of mJy. We then scaled the models by R$_*^2$/dist$^2$ using literature values for GJ 486: R$_*$=0.33 R$_\odot$ \citep{Trifonov2021} and dist = 8.07 pc \citep{Gaia2021}. We smoothed and interpolated the models to be the same resolution as the observations before calculating a reduced-$\chi^2$. In our reduced-$\chi^2$ calculations, we considered 3187 wavelength points for Transit 1 (3180 for Transit 2) and three fitted parameters ($T_{\rm eff}$, log($g$), and a scaling factor). The multi-component models included additional fit parameters for determining the percent coverage for the spots and faculae. The scaling factor was multiplied by the R$_*^2$/dist$^2$ term to account for uncertainty in either measured quantity and varied from 0.9 to 1.1. To get the final reduced-$\chi^2$ value for each model, we computed reduced-$\chi^2$ individually for Transits 1 and 2 and then took the average. 

Considering each type of one-, two-, and three-component model individually, we find that the models with the smallest reduced-$\chi^2$ values are fairly consistent with the existing literature values, though no model is a particularly good fit with a reduced-$\chi^2$ near 1 (for numerical details, see Appendix \ref{sec:bic}). A 100\% $T_{\rm eff}$ = 3300 K, log($g$)=4.5 cgs model with $\chi_\nu^2$=72.0 is the preferred one-component photosphere model (scale factor = 1.05), yielding a lower surface gravity than expected for a field age mid-M-dwarf like GJ~486. In agreement with our updated log($g$)=4.91$\pm$0.02 cgs, we disfavor the low stellar surface gravity of the best-matched photosphere-only model when taking into account inhomogeneities on the stellar surface. A 75\% $T_{\rm eff}$ = 3400 K, log($g$)=5 cgs background photosphere with 25\% spot coverage at $T_{\rm eff}$ = 3000 K, log($g$)=5 cgs is the preferred two-component model ($\chi_\nu^2$=53.4; scale factor = 1.05). The model most preferred overall is a three-component model with $\chi_\nu^2$=49.0 that has a background photosphere with $T_{\rm eff}$ = 3200 K, log($g$)=5 cgs, 20\% spot coverage at $T_{\rm eff}$ = 3000 K, log($g$)=5 cgs, and 25\% faculae coverage at $T_{\rm eff}$ = 3400 K, log($g$)=5 cgs (scale factor = 1.1). These three models are shown in Figure \ref{fig:stellar} compared to the Baseline GJ~486 spectra from Transits 1 and 2. There is decent general agreement for each model throughout the full $\sim$2.9-5 $\mu$m range, with slightly better agreement for the three-component photosphere+spot+faculae model, indicating that we cannot rule out star spots as a source for the presumed water detection.

\section{Discussion and Conclusions} \label{sec:conclusion}

There is remarkable agreement in the stellar heterogeneity parameters obtained from a) retrieving for unocculted star spots in the planetary transmission spectrum and b) fitting the baseline stellar spectrum with \texttt{PHOENIX} multi-component stellar models. Both lines of inquiry find best fits with overlapping values for faculae/spot coverage and temperature as well as the photospheric temperature. The stellar spectrum is best fit by a 3200 K photosphere with 20\% cool spots at 3000 K and 25\% hot faculae at 3400 K. These values match well compared to the TLS retrievals with a 3280 K photospheric temperature lower limit and cool spots up to $\sim$3100K at 7 - 18\% coverage
(see \Cref{fig:POSEIDON_summary}). This consistency lends strong support to this physical interpretation of our {\jwst} NIRSpec/G395H data. Moreover, even quiescent M dwarfs are known to be highly hetereogenous with strong impacts on the transmission spectrum \citep{Rackham2018,Zhang2018,Somers2020}.

Our forward model water atmosphere demonstrates that water is the best-fit absorber to explain \planetname's spectrum in the absence of stellar contamination. Such a pure steam atmosphere could theoretically be generated by impacts from small, icy bodies \citep{zahnle1988} or outgassed depending on the mantle composition \citep{TianHeng2023}, but would be quickly lost via the runaway greenhouse effect \citep{goldblatt2013}, as well as being disfavored by high resolution observations \citep{RiddenHarper2023}. We examine the effect of adding CO$_2$ to our H$_2$O forward model, finding that scaling the carbon content upwards always results in a worse fit to the data. In the water-rich \POSEIDON retrievals, we find strong water abundance lower limits across the three reductions, with an agnostic background gas prior. Two carbon species have stringent upper limits: 
carbon dioxide 
and methane. All reductions have posteriors where the constrained carbon species abundances can supersede that of water, but the best fits prefer atmospheres where water vapor dominates over carbon species. Such atmospheres would be challenging to maintain at \planetname's 700 K equilibrium temperature, given our current understanding of the runaway greenhouse effect \citep{goldblatt2013} and expected limits on the interior sequestration and outgassing rates of carbon species relative to water \citep{Sossi2023,TianHeng2023}. 
However, given the large range of retrieved abundances compatible with GJ 486's spectrum, they remain consistent with atmospheric theory. Furthermore, our retrievals cannot constrain the abundance of carbon monoxide (CO), providing an additional potential reservoir for carbon in the atmosphere. A warm, water-rich atmosphere with little atmospheric carbon would represent a terrestrial exoplanet wholly unlike any solar system analogue and challenge our understanding of atmospheric formation \citep{Wordsworth2022,McIntyre2023}.

\planetname joins the ranks of other terrestrial M-dwarf planets with tantalizing atmospheric inferences. Such planets include the first planet of our JWST-GO-1981 program, LHS~475b, exisiting observations of which cannot distinguish a carbon dioxide atmosphere from an airless body \citep{Lustig-YaegerFu2023}. L 98-59c is another planet where recent {\hst} observations have tentatively suggested either a hydrogen-rich planetary atmosphere or stellar contamination \citep{Barclay2023} --- though a different analysis favored a flat, featureless transmission spectrum \citep{Zhou2023}. Both \planetname at 1.3 R$_\oplus$ and L 98-59c at 1.35 R$_\oplus$ track the upper edge of planets below the expected hydrogen-dominated atmospheric cut-off \citep{Rogers2015,Rogers2021}. Their difference in insolation, with \planetname at T$_{\rm{eq}}$ = 700 K and L 98-59c at T$_{\rm{eq}}$ = 550 K, combined with their retrieved upper limit atmospheric hydrogen fractions, offer suggestive hints at a cosmic shoreline that is confounded by potential stellar contamination. More data are clearly necessary to confidently mark the boundaries of any cosmic shoreline.

Secondary eclipse observations of \planetname with {\jwst}'s  Mid-Infrared Instrument (MIRI) Low Resolution Spectroscopy (LRS) mode are already scheduled (GO 1743, PI: Mansfield). These observations will measure the dayside emission spectrum of the planet, allowing an expected $5\sigma$ constraint on surface pressures $\geq 1$ bar, as well as providing evidence for the atmospheric composition with a sufficiently thick atmosphere \citep{Mansfield2019,mansfieldjwst}. Thus, these MIRI/LRS observations can lend an additional line of evidence for or against both a significant atmosphere as well as the presence of water. However, our water-rich atmospheric retrieval scenario demonstrates that much lower surface pressures (down to millibar levels) are consistent with the data from NIRSpec/G395H, which is beyond the sensitivity of the planned MIRI/LRS observations. In this case, the secondary eclipse emission spectrum is unlikely to provide strong evidence in favor of either of our interpretations for \planetname.

As seen in Figure \ref{fig:retrieval}, the unocculted star spot scenario and the water-rich atmosphere scenario diverge strongly shortwards of 0.8 $\mu$m. In the case that the upcoming MIRI observations cannot definitely detect an atmosphere, high precision shorter wavelength observations could provide evidence for or against an atmosphere on \planetname. Ultimately, our {\jwst} NIRSpec/G395H stellar and transmission spectra, combined with retrievals and stellar models, suggest either an airless planet with a spotted host star or a significant planetary atmosphere containing water vapor. Given the agreement between our stellar modeling and atmospheric retrievals for the spot scenario, this interpretation may have a slight edge over a water-rich atmosphere. However, a true determination of the nature of \planetname remains on the horizon, with wider wavelength observations holding the key to this world's location along the cosmic shoreline. 

\section*{Acknowledgments}.
We thank the anonymous referee whose comments improved this manuscript. This work is based in part on observations made with the NASA/ESA/CSA \textit{JWST}. The data were obtained from the Mikulski Archive for Space Telescopes at the Space Telescope Science Institute, which is operated by the Association of Universities for Research in Astronomy, Inc., under NASA contract NAS 5-03127 for JWST.  These observations are associated with program \#1981.
Support for program \#1981 was provided by NASA through a grant from the Space Telescope Science Institute, which is operated by the Association of Universities for Research in Astronomy, Inc., under NASA contract NAS 5-03127. 
This material is based in part upon work performed as part of the CHAMPs (Consortium on Habitability and Atmospheres of M-dwarf Planets) team, supported by the National Aeronautics and Space Administration (NASA) under Grant No. 80NSSC21K0905 issued through the Interdisciplinary Consortia for Astrobiology Research (ICAR) program. 
The material is based upon work supported by NASA under award number 80GSFC21M0002.
We also acknowledge Jordin Sparks for her lyrical genius.

\vspace{5mm}
\facilities{JWST(NIRSpec)}
All the {\it JWST} data used in this paper can be found in MAST: \dataset[10.17909/z89v-dg97]{http://dx.doi.org/10.17909/z89v-dg97}.

\software{ Astropy \citep{astropy,astropy2}, \texttt{batman} \citep{batman2015}, \texttt{CHIMERA} \citep{Line2013b,Line2014-C/O}, \texttt{Dynesty} \citep{Speagle2020}, \texttt{emcee} \citep{emcee2013}, \eureka \citep{Eureka2022}, ExoCTK \citep{exoctk}, \texttt{FIREFLy} \citep{Rustamkulov2022}, \texttt{Forecaster} \citep{Chen2017}, IPython \citep{ipython}, \texttt{jwst} \citep{jwstpipeline2022}, Matplotlib \citep{matplotlib}, NumPy \citep{numpy, numpynew}, \texttt{PHOENIX} \citep{Allard2012} \texttt{PICASO} \citep{Batalha2019}, \texttt{POSEIDON} \citep{MacDonald2017,MacDonald2023}, PyMC3 \citep{Salvatier2016}, pysynphot\citep{STScIDevelopmentTeam2013}, \rfast \citep{RobinsonArnaud2023}, SciPy \citep{scipy}, 
\texttt{Tiberius} \citep{Kirk2019,Kirk2021}}



\newpage
\appendix

\section{Data Reduction} 
\label{sec:data}

\begin{table}[h!]
    \centering
    \begin{tabular}{c|c|c|c|c|c} \hline
Dataset         & $T_0$ (BJD$_{TDB}$)                   & $i$ ($^{\circ}$)          & $a/Rs_*$                  & $R_P/R_*$     & Residual RMS (ppm) \\ 
\hline
Transit 1, NRS1 & $2459939.071619^{+2.0e-05}_{-2.1e-05}$  & $89.10^{+0.26}_{-0.35}$   & $11.24^{+0.03}_{-0.09}$   & $0.03697 \pm 0.00009$ & 143 \\
Transit 1, NRS2 & $2459939.071570^{+2.4e-05}_{-2.4e-05}$  & $89.06^{+0.34}_{-0.38}$   & $11.22^{+0.06}_{-0.13}$   & $0.03784 \pm 0.00009$ & 171 \\
Transit 2, NRS1 & $2459943.472959^{+2.0e-05}_{-2.0e-05}$  & $89.02^{+0.35}_{-0.38}$   & $11.23^{+0.07}_{-0.13}$   & $0.03689 \pm 0.00009$ & 137 \\
Transit 2, NRS2 & $2459943.472974^{+2.3e-05}_{-2.4e-05}$  & $89.06^{+0.46}_{-0.47}$   & $11.22^{+0.10}_{-0.19}$   & $0.03670 \pm 0.00009$ & 158 \\ 
\hline
Weighted Mean & $2459939.071594 \pm 1.6e-05$            & $89.06 \pm 0.18$          & $11.229 \pm 0.043$       & $0.03709 \pm 0.00004$   & n/a \\ 
              & $2459943.472967 \pm 1.5e-05$ & & & &\\
\hline
    \end{tabular}
    \caption{Best-fit system parameters and 1$\sigma$ uncertainties from fitting the four white light curves using \eureka.}
    \label{tab:Eureka_sys_params}
\end{table}

\begin{table}[h!]
    \centering
    \begin{tabular}{c|c|c|c|c|c} \hline
Dataset & $T_0$ (BJD$_{TDB}$) & $i$ ($^{\circ}$) & $a/Rs_*$ & $R_P/R_*$ & Residual RMS (ppm) \\ \hline
Transit 1, NRS1 & 2459939.0716102 $\pm$ 2.1e-05 & 89.11 $\pm$ 0.35 & 11.294 $\pm$ 0.137   & 0.03759 $\pm$ 0.00013&  132\\
Transit 1, NRS2 & 2459939.0715592 $\pm$ 2.2e-05 & 89.97 $\pm$ 0.27 & 11.449 $\pm$ 0.023   & 0.03791 $\pm$ 0.00010 & 159 \\
Transit 2, NRS1 & 2459943.4729689 $\pm$ 1.9e-05 & 89.99 $\pm$ 0.22 & 11.446 $\pm$ 0.021  & 0.03784 $\pm$ 0.00013 & 130 \\
Transit 2, NRS2 & 2459943.4730019 $\pm$ 2.3e-05 & 89.30 $\pm$ 0.40 & 11.325 $\pm$ 0.111   & 0.03742 $\pm$ 0.00017 & 158 \\ \hline
Weighted Mean & 2459939.0715859 $\pm$1.5e-05 & 89.75 $\pm$ 0.14 & 11.443 $\pm$ 0.015 & 0.03775 $\pm$ 0.000063 & n/a\\ 
& 2459943.4729823 $\pm$ 1.5e-05 &  &  & & \\ \hline
    \end{tabular}
    \caption{The system parameters resulting from the \firefly fits to the white light curves. }
    \label{tab:firefly_sys_params}
\end{table}


\begin{table}[h!]
    \centering
    \begin{tabular}{c|c|c|c|c|c} \hline
Dataset & $T_0$ (BJD$_{TDB}$) & $i$ ($^{\circ}$) & $a/Rs_*$ & $R_P/R_*$ & Residual RMS (ppm) \\ \hline
Transit 1, NRS1 & $2459939.071586^{+3.5e-05}_{-3.6e-05}$ & $89.99^{+0.65}_{-0.61}$ & $11.34^{+0.04}_{-0.13}$ & $0.03683 \pm 0.00015$ & 158 \\
Transit 1, NRS2 & $2459939.071548^{+3.6e-05}_{-3.5e-05}$ & $89.97^{+0.67}_{-0.62}$ & $11.36^{+0.05}_{-0.13}$ & $0.03756 \pm 0.00017$ & 188 \\
Transit 2, NRS1 & $2459943.472952^{+3.6e-05}_{-3.5e-05}$ & $90.02^{+0.75}_{-0.72}$ & $11.42^{+0.06}_{-0.17}$ & $0.03684 \pm 0.00015$ & 158 \\
Transit 2, NRS2 & $2459943.472955^{+4.3e-05}_{-4.3e-05}$ & $89.83^{+1.35}_{-1.32}$ & $11.23^{+0.19}_{-0.4}$ & $0.03685 \pm 0.00019$ & 194 \\ \hline
Weighted Mean & $2459939.07158 \pm 1.9e-05$ & $89.96 \pm 0.37$ & $11.40 \pm 0.06$ & $0.03701 \pm 8e-05$ & n/a \\ \hline
    \end{tabular}
    \caption{The system parameters resulting from the \texttt{Tiberius} fits to the white light curves.}
    \label{tab:Tiberius_sys_params}
\end{table}

\subsection{Data Reduction Consistency: An Offset between the NRS1 and NRS2 Detectors} \label{sec:offset}
As stated in the main text, all initial reductions showed a consistent offset in measured transit depth for the Transit 1, NRS2 detector relative to the other white light curve depths. Since this shift is not seen in the NRS1 detector, we can confidently rule out all astrophysical effects (e.g., stellar variability) as a source of the discrepancy. For the \firefly reduction, we altered our application of the superbias in the bias subtraction step and light-curve fitting stages, which we found produced more consistent transit depths for NRS1 and NRS2.

In our \firefly reduction, we measured the superbias level by rescaling the superbias image to match the level in the trace-masked groups of each integration. We note that a full readout of the detector mitigates bias drifts using reference pixels, but the subarray readouts used here do not have such pixels. We find that the superbias level changes by hundreds of ppm throughout the time series, with typical values of the scaling factor about 1.003. We use the standard-deviation-normalized time series of the superbias scaling coefficient as a detrending vector at the light-curve fitting stage, added linearly to our usual systematics model. We find that the superbias decorrelation coefficient is statistically preferred in the systematics model, with some residual structure in the photometry well-explained by this term. The addition of superbias detrending reduced the transit depth tension between NRS1 and NRS2, with the white-light curve transit depths agreeing within the uncertainties.

For the \eureka reduction, we also investigated time-dependent variations in the NRS2 detector bias level.  We found that applying a scale factor correction to the superbias frame for each integration in Stage 1 marginally improved the consistency in measured transit depths (by $\sim20$ ppm), but also led to increased scatter.  Applying a single scale factor correction for all integrations yielded a similar improvement, but without the increased scatter.  We continue to investigate different methods of scaling the superbias frame.  In the meantime, we elect to adopt the standard bias correction in our final {\eureka} analysis and apply a manual offset of 78 ppm in transit depth to NRS2, Transit 1. 

To account for NRS2 transit visit discrepancy for the final \tiberius reduction, we also manually offset the transmission spectrum for NRS2, Transit 1 by 63 ppm, such that the median transit depth was equal to NRS2, Transit 2.

After this superbias-detrending in \firefly and manual offsets in \eureka and \tiberius, we saw excellent agreement between the \eureka, \firefly, and \tiberius spectra across both NRS1 and NRS2 in both transits, as shown in \Cref{fig:gaussian_spectra}.  Since the superbias correction alters \firefly's absolute transit depths, we elect to compare their relative transit depths.

\section{Interpretation Supplemental Information}
\label{sec:forward_and_retrieval}

\begin{table}[h]
\centering
\begin{tabular}{c|c|c|c|c|c}
\hline
\textbf{\texttt{CHIMERA}} &
  \eureka   & \firefly & \tiberius & Average $\sigma$ & Significance \\ 
 \textbf{Forward Model} & (dof = 110) & (dof = 46) & (dof = 46) & ruled out \\ \hline
1000$\times$ solar & 1.64 & 1.26 & 2.44 & 3.6 & moderately ruled out \\ 
H$_2$O, 1 bar      & 1.01 & 0.76 & 1.37 & 0.9   & consistent with data  \\ 
CO$_2$, 1 bar      & 1.39 & 1.17 & 1.63 & 2.3 & weakly/moderately ruled out \\ 
CH$_4$, 1 bar      & 2.10 & 1.77 & 5.96 & 6.5 & strongly ruled out \\ 
Earth-like         & 1.33 & 1.04 & 2.35 & 2.8 & moderately ruled out \\ 
Flat line      & 1.11 & 0.91 & 1.60 & 1.5 & weakly/moderately rejected by Gaussian fitting \\ \hline
\end{tabular}
   \caption{Each reduction's reduced-$\chi^2$ compared to our end-member composition \picaso forward models. Since each reduction has a different degree-of-freedom (dof), we also report the average significance (in $\sigma$, following \citet{Trotta2017}) by which the model is ruled out. Note that the ``flat line'' model can correspond either to an airless planet or a very hazy atmosphere.}
    \label{tab:models}
\end{table}

\begin{figure*}[h!]
    \centering
    \includegraphics[width=\textwidth]{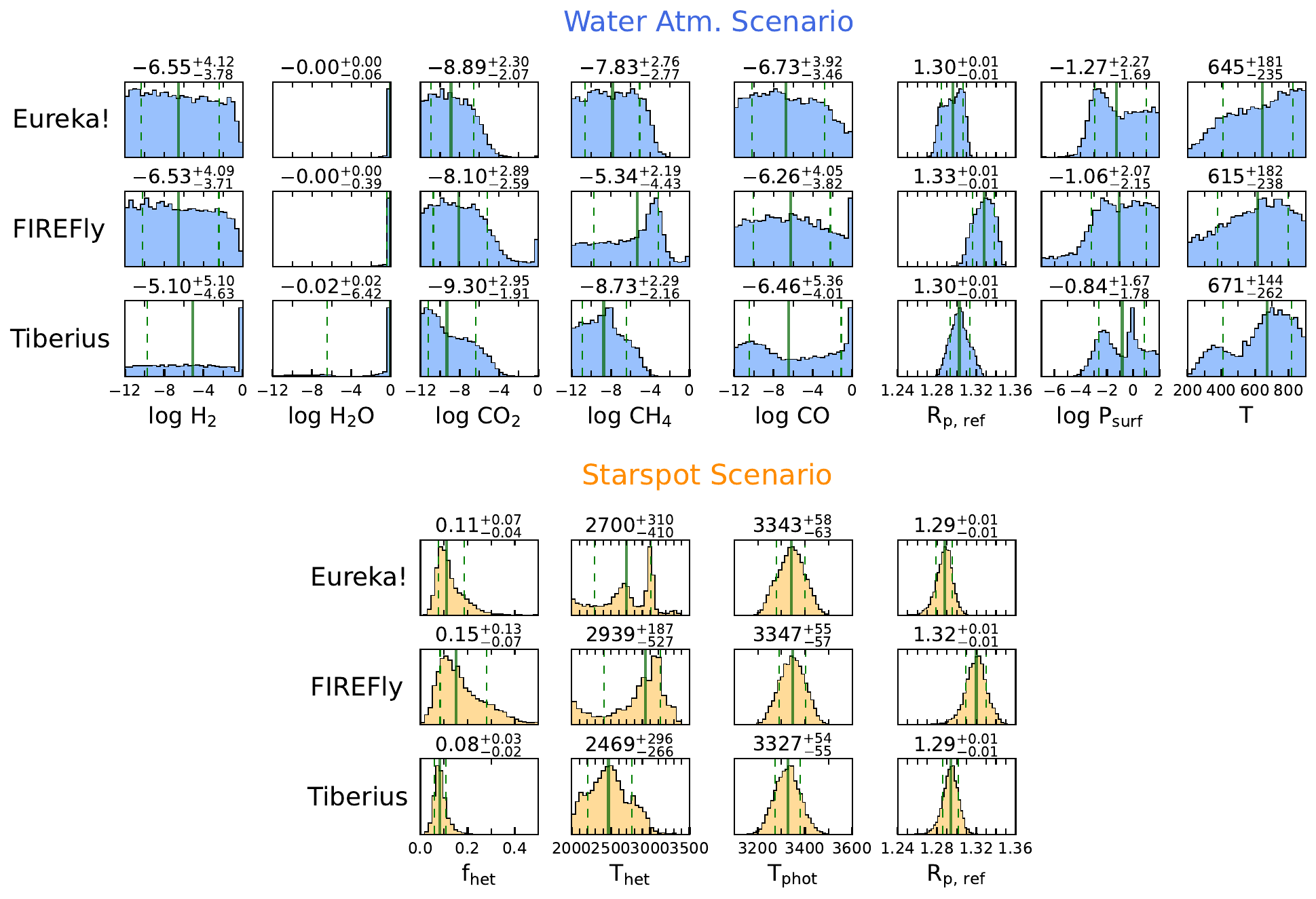}
    \caption{Posterior probability distributions from the \POSEIDON retrievals. Top rows (blue): retrieval model where \planetname's spectrum is caused by a water-rich atmosphere. Bottom rows (orange): retrieval model instead considering unocculted starspots. The rows in each scenario correspond to different data reductions (\eureka, \firefly, and \tiberius from top to bottom). 
    }
    \label{fig:POSEIDON_summary}
\end{figure*}

\begin{figure}[h!]
    \centering
    \includegraphics[width=0.9\linewidth]{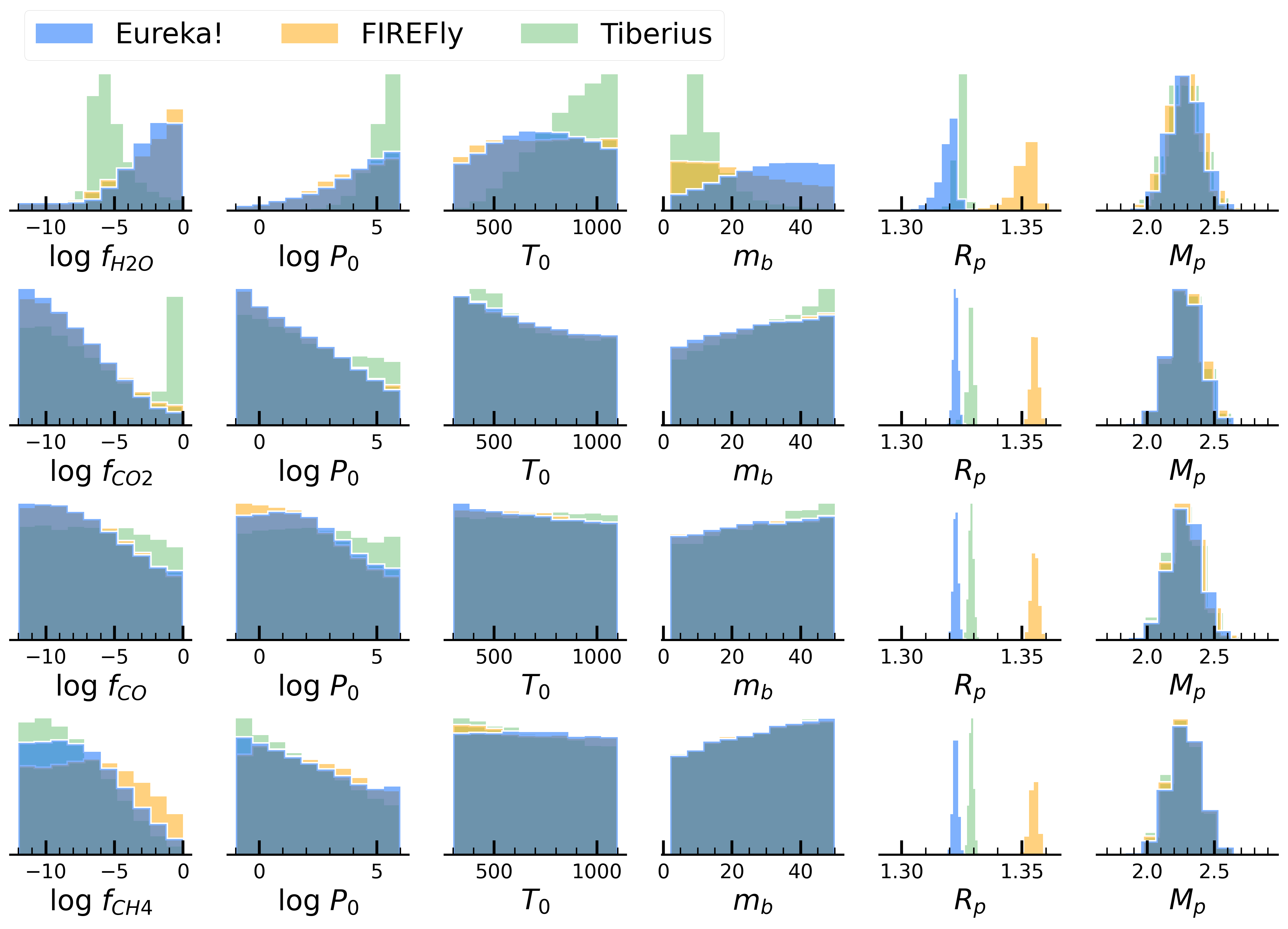}
    \caption{The 1D posteriors for \rfast single-gas retrievals. From top to bottom water, carbon dioxide, carbon monoxide, and methane. Each reduction is shown in its own color with \eureka in blue, \firefly in orange and \tiberius in green. 
    }
    \label{fig:rfast_summary}
\end{figure}

\section{Stellar Model Statistics}
\label{sec:bic}

\begin{table}[h]
\centering
\begin{tabular}{c|c|c|c|c}
\hline
Model Configuration &
  $\chi_\nu^2$   & $\chi^2$ & $n$ & $K$\\ \hline
 Photosphere               & 72.0 & 228,680.81 & 3,187 & 3 \\ 
 Photosphere+Spot          & 53.4 & 169,489.64 & 3,187 & 5\\
 Photosphere+Spot+Faculae  & 49.0 & 155,374.94 & 3,187 & 7\\\hline
\end{tabular}
   \caption{Summary of values for our goodness-of-fit testing, where $n$ is the number of wavelength points, and $K$ is the number of free parameters.}
    \label{tab:bic}
\end{table}


\newpage
\bibliography{NoAir486b}{}
\bibliographystyle{aasjournal}



\end{document}